# Free-space quasi-phase matching

NAZAR KOVALENKO,[1,*] VICTOR HARITON[1], KILIAN FRITSCH[1] AND OLEG PRONIN[1]

[1]*Helmut Schmidt University, Holstenhofweg 85, 22043 Hamburg, Germany*
*Corresponding author: nazar.kovalenko@hsu-hh.de*

**We report a new approach to phase matching of nonlinear materials based on the free space multipass cells. This concept quasi-phase matches crystalline quartz and increases the second harmonic generation efficiency by a factor 40.**

Nonlinear optical effects based on birefringent phase matching and quasi-phase-matching (QPM) are an inherent part of contemporary photonics. These effects made it possible to design and create unique spatial and temporal conditions for nonlinear interactions. They enabled many applications by providing novel sources of broadband and narrowband laser radiation [1].

Well-established available technology for producing periodically poled structures applies only to ferroelectrics – crystalline materials with a spontaneous electric polarization, reversible by applying an external electric field. Among the widely used nonlinear optical materials, numbering more than 80 compounds, only three (KTP, LN, LT) are suitable for creating commercially available periodically poled (PP) structures and, unfortunately, very limited in their transparency range 340-5500 nm. Different approaches were used to create conditions for QPM in other materials. For example, in [2], and [3] it was realized in Brewster-angle-stacked wafers (5 plates), made of GaAs and CdTe materials. In [4] QPM conditions were created in chalcogenide crystals by growing these crystals with specific packing defects of the crystalline structure. Attempts were made to create a tight stack of multiple, bonded-together crystalline plates [5]. Due to the rather low practicability and high complexity of these approaches they did not receive further development.

Here we propose a concept based on multiple passes of interacting beams through the same nonlinear crystal (NL) element placed inside a quasi-waveguide or multi-pass cell (see Fig.1b).

The operational principle of this free-space quasi-phase matching (FSQPM) scheme can be well explained using a comparison with the well-known QPM scheme in periodically poled structures [6]. In these structures, the relative phase between the interacting waves can be maintained by periodically reversing the sign of the nonlinear susceptibility and this way keeping the buildup of the second harmonic output $E_{2\omega}$ (see Fig. 1a). However, in the FSQPM scheme, the build-up of the second harmonic from pass to pass is achieved by tuning the relative phase with a help of medium with negligible nonlinearity and reasonable linear dispersion (fused silica (FS) plates in Fig. 1b). In a quasi-waveguide geometry, after each pass through the nonlinear element (NL) this correction of the relative phase ψ takes place in the medium FS (Fig.1b) and then the beam gets refocused in the NL again.

This type of FSQPM can be realized for all types of nonlinear materials increasing the length of the nonlinear interaction and maximizing intensity in the crystal. Moreover, birefringent phase matching can also be realized in multipass cells leading to drastically reduced pump intensities in the nonlinear crystals.

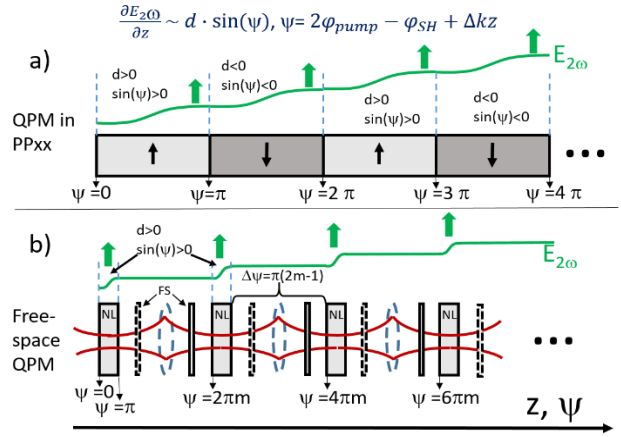

Fig.1 Comparing QPM conditions in periodically poled structures and free-space QPM in a multipass cell: $E_{2\omega}$-amplitude of the second harmonic electric field, $\frac{\partial E_{2\omega}}{\partial z}$-first derivative of the $E_{2\omega}$, ψ-relative phase, $\varphi_{pump}$ – phase of the pumping field, $\varphi_{SH}$ – phase of the second harmonic, Δk – wave number mismatch, m-integer, NL-nonlinear element, FS – fused silica phase corrector. The free space QPM is schematically represented as a quasi-waveguide formed by the lenses.

The experimental setup is shown in Fig.2 containing two concave 1-inch diameter mirrors with radii of curvature of 500 mm separated by 950 mm distance. The nonlinear crystal NL is placed in the center of the Herriot-type multipass cell. The relative phase correction is performed by two anti-reflected coated plates of fused silica with a thickness of 3 mm. An Nd:YAG laser operating at 1.064 μm in Q-switched mode with a 6 ns pulse duration and repetition rate of 10 Hz is used as a pump source. Firstly, we performed free space QPM experiments with crystalline quartz. This material is available in excellent optical quality, has a unique transparency range (140-3500 nm), and, interestingly, was the first material to be used for the seminal work of Franken [7] on the second harmonic generation. Unfortunately, this crystal with its coherence length of 21 μm for 1064 nm wavelength cannot be birefringently phase-matched. At pump intensity of 500 MW/cm² and for 14 passes through an uncoated crystalline quartz nonlinear element, a 12-fold increase in the second harmonic beam intensity was obtained compared to a single pass. The crystal quartz plate had no antireflection coating showing the reflection losses of 8.3% for a single pass which corresponds to a 40-fold increase in the second harmonic when taking these losses into account. The beam profile of the generated green radiation at 532 nm showed a perfect Gaussian shape after 14 passes, similar to the input beam profile at 1064 nm (see Fig.2). In the second experiment, we used a pair of KTP crystals, each 1 mm thick and 10x10 mm² in size, to show that the concept is applicable to the birefringently phase-matched nonlinear crystals.

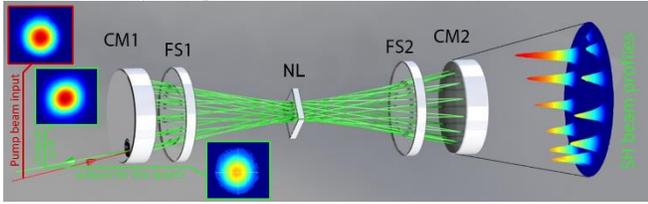

Fig.2 The multipass cell with a nonlinear crystal inside: CM1, CM2- curved mirrors, NL-nonlinear crystal, FS1, FS2-fused silica plates. On the right side, the build-up of second harmonic generation monitored by a CCD camera behind the CM2 mirror is shown. One can see that the signal grows for every next pass through the NL material. NL material in our setup is crystalline quartz or a pair of KTP crystals.

The crystals were oriented to compensate for the spatial walk-off effect. Additionally, one of the phase-tuning plates FS2 was removed, and the desired phase shift was obtained in ambient air by changing the distance between KTP crystal and mirrors CM2. The cell was increased to 975 mm to adopt 20 passes through the nonlinear crystals. The spot diameter in the nonlinear crystal was about 320 µm corresponding to 73 MW/cm² peak intensity in the crystal and 0.44 J/cm² fluence on the mirrors. This resulted in a conversion efficiency of ~63% (see Fig 3) which is comparable to conventional 1-pass conversion efficiency obtained in a KTP crystal with a thickness of 10 mm [8] at higher intensities of >80 MW/cm². Such relatively low pump intensity implies a longer lifetime for the nonlinear element, particularly in the UV range, where the crystal degradation is significant. The demonstrated increase in the effective length of the interaction of radiation makes it possible to fully reveal the potential of different nonlinear crystals, including those which cannot be grown in sufficient thickness or are prone to damage such as DAST or KBBF crystals.

The possibility of realizing quasi-phase-matching in crystalline quartz, transparent down to 140 nm, opens the prospect of using this material for generating vacuum ultraviolet at around 150 nm wavelength for Thorium nuclear clocks. For example, by taking the equations from [3] and applying parameters of crystalline quartz, we can obtain estimated conversion efficiency from 355 nm to 177.5 nm of 0.1% for the pump power to 100 GW/cm$^{-2}$ and 100 passes in the Herriot cell.

The Herriot cells with such a number of passes can be easily realized with 2-inch diameter mirrors. The ability to change the beam parameters before each pass through the nonlinear element also makes it possible to use the proposed scheme for converting ultrashort pulses and compensating the dispersion after each pass by means of dispersive mirrors. The proposed approach applies to all types of nonlinear crystals. Thus, in optically isotropic nonlinear materials, there is no need to adhere to special directions of propagation of interacting radiations required for the classic phase-matching. The conversion efficiency scales with $N^2$ where N is the number of passes through the nonlinear crystals [2]. Thus, it can be readily increased by about four orders of magnitude for the multipass cells implementing about 100 passes. Additionally, the nonlinear element can be kept within the Rayleigh length of the focused pump radiation inside the cell and thus maintain a relatively homogeneous and if necessary high pump radiation intensity at each pass through the nonlinear element during the parametric conversion process.

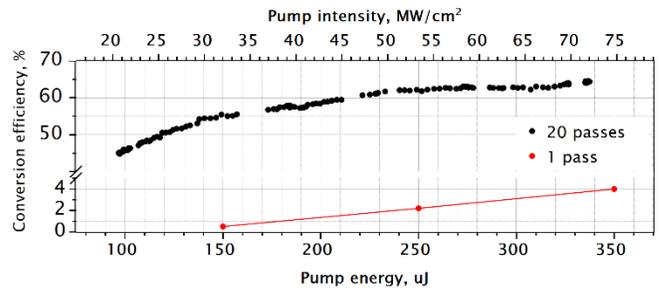

Fig. 3 Conversion efficiency of the pump radiation into the second harmonic vs pump energy and intensity. The red curve corresponds to a single pass and the black curve to 20 passes inside a multipass cell.

This method is also applicable in combination with random quasi-phase matching in a simplified scheme without using the phase tuning elements. A promising option is the ability to tune the phase shift by varying the parameters of the gas pressure and temperature. This provides extremely precise and continuous phase shift for all beam passes inside the multipass cell.

In summary, $\chi^{(3)}$ based nonlinear effects were extensively demonstrated in multipass cells over the last 6 years providing and establishing the field of $\chi^{(3)}$ multipass nonlinear optics [9]. This work shows the first proof-of-concept demonstration of $\chi^{(2)}$ multipass nonlinear optics where free space birefringent phase matching and quasi-phase matching can be realized with nearly all types of nonlinear materials.